# Relaxation of Shallow Donor Electron Spin due to Interaction with Nuclear Spin Bath

*Semion Saykin,[1,2] Dima Mozyrsky,[3] and Vladimir Privman[1,\*]*

[1] Center for Quantum Device Technology, Clarkson University, Potsdam, New York 13699, USA

[2] Department of Theoretical Physics, Kazan State University, Kazan 420008, Russia

[3] T-13 and CNLS, Los Alamos National Laboratory, Los Alamos, New Mexico 87545, USA

[\*] E-mail: privman@clarkson.edu

**Abstract.** We study the low-temperature dynamics of a shallow donor, e.g., [31]P, impurity electron spin in silicon, interacting with the bath of nuclear spins of the [29]Si isotope. For small applied magnetic fields, the electron spin relaxation is controlled by the steady state distribution of the nuclear spins. We calculate the relaxation times $T_1$ and $T_2$ as functions of the external magnetic field, and conclude that nuclear spins play an important role in the donor electron spin decoherence in Si:P at low magnetic fields.

**Introduction.** Recently, there has been much interest in the studies of decoherence of a single electron or nuclear spin for novel low temperature semiconductor applications considered for realization of quantum information processing. Several proposed designs of quantum computers[1-7] utilize spin qubits. Interactions with environment lead to deviations from controlled, coherent quantum-mechanical evolution of the spin state. Traditionally, the loss of the initial spin polarization, has been characterized by the longitudinal, $T_1$, and transverse, $T_2$, relative to the direction of the external magnetic field, relaxation times.[8] Interest in quantum computing has focused attention on the time scales of the processes involved and on the properties of single spins. Much of the recent work has been devoted to the finding[9-15] that initial decoherence processes may be important for quantum computing. These processes occur on time scales faster than energy exchange. In this work, we focus on a system suggested by quantum computing designs, but concentrate on the global time dependence of decoherence and relaxation.

A spin-qubit in a semiconductor heterostructure at low temperatures, can be nuclear spin,[2,3] or spin of an electron bound to a donor impurity[5] or trapped in a quantum dot.[1,4,6] This spin interacts with several types of environment, such as phonons, conduction electrons, and other spins. It has been argued in the literature that the spin environment possesses fundamentally different properties from the bosonic one.[12,13,16,17] In the present work, we study effects of the nuclear spin bath on the impurity-bound electron spin qubit.

It is well established that the nuclear spin system can influence electron spin polarization. This has been demonstrated for GaAs, where polarized nuclei can create strong (of order several Tesla) effective internal magnetic field at an electron position.[18] The nuclear spin system has long relaxation time as compared to the electron spin system, and for short times can be considered as system of "frozen" magnetic moments, unless the nuclear spins are pumped externally by NMR radiation. Recently, electron spin dephasing by nuclear spins in GaAs quantum dots was studied in Ref. 19, where the case of fully polarized nuclear spins was solved exactly and nonexponential decoherence processes were found. A different mechanism for irreversible spin-flip transitions with energy dissipation in quantum dots due to the hyperfine interaction assisted by phonons, was proposed in Ref. 20-21, but it cannot lead to significant decoherence and relaxation. A more effective phonon-mediated decoherence mechanism[22] is due to spin-orbit interactions, as referenced later. Nuclear-spin driven dephasing of electron spins in GaAs quantum dots was considered recently[23] within a model of electron spins moving in effective magnetic fields created by contact hyperfine interactions.

In this paper, we consider relaxation of a shallow donor impurity, $^{31}$P ($I_P = 1/2$), electron spin in a defect-free Si crystal at low external magnetic fields and low temperature. The interest to this system has been rekindled by the work of Kane,[3] in which donor electron spins were proposed as mediators of interactions between nuclear spin qubits. We propose a novel mechanism for localized electron spin relaxation/decoherence in Si:P, that arises due to the precession of the electronic spin in a randomly distributed nuclear spin effective magnetic field. The nuclear spin bath couples to the electron spin via the hyperfine interaction. A well-known dephasing mechanism was suggested in Ref. 24, with electron spin dephasing by the nuclear spin polarization occurring as a result of hopping of the electron from one donor site to another. In this work, we establish that nuclear spins contribute to localized electron spin dephasing even in crystals with sparsely positioned donors, where such hopping is not possible. Our results indicate that nuclear spins are likely to be an important source of the low-temperature donor electron spin relaxation/dephasing in Si:P at low magnetic fields.

**Donor Electron Spin Interacting with a Nuclear Spin Bath.** In the effective mass approximation, donor electron in Si is described[25] by the linear combination of modulated *s*-electron wave functions,

$$\psi(\mathbf{r}) = \sum_{i=1}^{6} \alpha_i F_i(\mathbf{r}) u_i(\mathbf{r}) e^{i\mathbf{k}_i \cdot \mathbf{r}}, \qquad (1)$$

where the envelope function is $F_i(\mathbf{r}) = (\pi b^3)^{-1/2} \exp(-r/b)$. The effective radius is $b = \varepsilon R_b m/m^* \approx 20-25\,\text{Å}$, where $\varepsilon = 12$ is the dielectric constant for silicon, $R_b$ is the Bohr radius, while $m$ and $m^*$ are the free and effective electron masses, respectively. In Eq. 1, coefficients $\alpha_i$ are determined by the symmetry considerations, and the summation is carried over the six minima of the conduction band.[25] Natural silicon crystal contains about $c \approx 4.67\%$ of $^{29}$Si atoms with nuclear spin $I = 1/2$. The lattice constant is $a = 5.43$ Å, and thus the electron wave function with such effective radius covers $n_1 \approx 80$ nuclei of $^{29}$Si. We will assume that the electron spin interacts with the donor nucleus and nuclear spin bath of $^{29}$Si mainly via the contact hyperfine interaction.[8]

At low temperatures, of order several 10 mK, the donor electrons are always bound, and the role of phonons in relaxation processes is diminished. We can then focus on the spin Hamiltonian for a donor electron spin, **S**, interacting with a reservoir of nuclear spins, $\mathbf{I}^i$, in external magnetic field, **H**. It can be approximated by the Zeeman energy of the electron and nuclear spins, and by the hyperfine contact interaction

$$H = g\mu_b \mathbf{S}\mathbf{H} + \sum_i \gamma_i \hbar \mathbf{I}^i \mathbf{H} + \sum_i A_i \hbar \mathbf{S}\mathbf{I}^i, \qquad (2)$$

where summation is carried over all the nuclear spins. The spin-orbit interaction mixes ground and excited donor electron states described by Eq. 1, and can give rise to spin-lattice relaxation[26] and decoherence[22] by modifying the electron $g$-factor. This is important for phonon-mediated relaxation mechanisms.[22,26] In our calculation, it can be assumed that the electronic $g$-factor is isotropic and equal to 2. In Eq. 2, the hyperfine coupling constant of the $i$'s nuclear spin to the electronic spin is $A_i = (8\pi/3)\hbar g \mu_b \mu_n^i |\psi(\mathbf{r}_i)|^2$, where $\mu_n^i$ is magnetic moment of the nucleus located at position $\mathbf{r}_i$.

Our results, to be presented shortly, suggest that the main relaxation effects due to the nuclear spin bath occur in the regime when the effective magnetic field owing to the hyperfine interaction approximately cancels the external applied magnetic field. Therefore, we will focus on the magnetic fields of less than order 100 G, so that the nuclear Zeeman part of the Hamiltonian Eq. 2 is much smaller than the hyperfine part and can be neglected. Then the Hamiltonian Eq. 2 can be transformed to the interaction picture, $H_{int}(t) = \exp(iH_0 t) H \exp(-iH_0 t)$, with

$$H_0 = g\mu_b S_z \mathrm{H} + \sum_i A_i S_z I_z^i, \tag{3}$$

where we have included the diagonal part of the hyperfine interaction in the unperturbed Hamiltonian $H_0$.

The interaction Hamiltonian $H_{int}(t)$ can be split into two parts,

$$H_{int}(t) = H_{Si}(t) + H_P(t), \tag{4}$$

where

$$H_{Si}(t) = 2\sum_{i\neq 0} A_i \left\{ S_+ I_-^i e^{i\hat{\omega}_{Si}^i t} + S_- I_+^i e^{-i\hat{\omega}_{Si}^i t} \right\}, \tag{5}$$

$$H_P(t) = 2A_0 \left\{ S_+ I_-^0 e^{i\hat{\omega}_P t} + S_- I_+^0 e^{-i\hat{\omega}_P t} \right\}. \tag{6}$$

In the above equations, $H_{Si}$ and $H_P$ represent the off-diagonal part of the hyperfine interaction with the $^{29}$Si nuclei and P nucleus, respectively. It should be noted that the "frequencies"

$$\hat{\omega}_P = \omega_Z + \sum_j A_j I_z^j, \tag{7}$$

$$\hat{\omega}_{Si}^i = \omega_Z + A_0 I_z^0 + \sum_j{'} A_j I_z^j, \tag{8}$$

are defined as operators in nuclear spin space, and $\omega_Z = g\mu_b \mathrm{H}/\hbar$. The prime in the sum in Eq. 8 indicates that the summation is over $j \neq i$. The $j = i$ term can be included with negligible error, assuming a large number of spins in the reservoir. The energy level structure of the unperturbed Hamiltonian $H_0$ is schematically presented in Fig. 1. The hyperfine splitting produced by the donor nucleus is much larger than the splitting due to the interaction with the $^{29}$Si nuclei, as determined in the ENDOR experiments[27] for Si:P, specifically, $\Delta H_P \approx 42$ G and $\delta_{Si} \approx 2.9$ G, see Fig. 1.

In order to evaluate dynamics of the system governed by the Hamiltonian Eq. 4, we use a Markovian approximation for the master equation for the reduced density matrix $\rho(t)$ of the donor electron spin,[28]

$$\dot{\rho}(t) = -\int_0^t d\tau \{ [S_+, S_- \rho(t)] \xi_1(\tau) + [\rho(t) S_-, S_+] \xi_2(\tau) \qquad (9)$$
$$+ [\rho(t) S_+, S_-] \xi_1(-\tau) + [S_-, S_+ \rho(t)] \xi_2(-\tau) \},$$

where

$$\xi_1(\tau) = \sum_{i \neq 0} (2A_i)^2 \langle I_-^i I_+^i \rangle \langle e^{i\hat{\omega}_{S_i}\tau} \rangle + (2A_0)^2 \langle I_-^0 I_+^0 \rangle \langle e^{i\hat{\omega}_P \tau} \rangle, \qquad (10)$$

$$\xi_2(\tau) = \sum_{i \neq 0} (2A_i)^2 \langle I_+^i I_-^i \rangle \langle e^{i\hat{\omega}_{S_i}\tau} \rangle + (2A_0)^2 \langle I_+^0 I_-^0 \rangle \langle e^{i\hat{\omega}_P \tau} \rangle. \qquad (11)$$

In the above equation, the angular brackets denote averages over the spin states, obtained by tracing the appropriate operators multiplied by the density matrix of the spin bath, $\theta(\tau)$. We point out that the approximations involved[28] in deriving Eq. 9 include the assumption that the total density matrix is factorized at all times. Furthermore, the density matrix of the bath is assumed to be time-independent. These assumptions of the Markovian approximation, are generally valid when the nuclear spin reservoir is kept in its reference state either by external pumping by NMR radiation, or by thermalization. These processes, as well as interactions present in the system, will define the time scales of the decay of $\xi_{1,2}(\tau)$, which should be smaller than the characteristic dynamical times of the electron spin, $T_1$ and $T_2$. Thermalization processes alone might not be sufficient to satisfy this condition for experimentally relevant times. This limitation should be kept in mind when using the results of most recently published relaxation calculations mediated by nuclear spins.[19,23]

In our calculations, we took the completely random $\theta = 2^{-n_I}$, assuming that any experimentally relevant temperature is effectively infinite for nuclear spins, or that they are continuously pumped. Averaging of exponential operators,

$$\langle \exp(i\hat{\omega}_{S_i}\tau) \rangle = \exp(i\omega_Z \tau) \langle \exp(iA_0 I_z^0 \tau) \rangle \langle \exp(i \sum_j A_j I_z^j \tau) \rangle, \qquad (12)$$

in Eqs. 10, 11 can be done assuming that the number of nuclear spins is sufficiently large,

$$\left\langle e^{\pm i \sum_j A_j I_z^j \tau} \right\rangle \approx \frac{1}{\sqrt{2\pi\sigma^2}} \int_{-\infty}^{\infty} dy\, e^{-\frac{y^2}{2\sigma^2} \pm iy\tau} = e^{-\tau^2\sigma^2/2}, \qquad (13)$$

where the energy scale

$$\hbar\sigma = \hbar\sqrt{\sum_j (A_j/2)^2} \qquad (14)$$

measures the root-mean-square hyperfine interaction of the electron with the nuclear bath. Thus,

$$\left\langle e^{i\hat{\omega}_S \tau} \right\rangle = e^{-\tau^2\sigma^2/2 + i\omega_Z \tau} \cos(A_0\tau/2), \qquad (15)$$

and

$$\left\langle e^{i\hat{\omega}_P \tau} \right\rangle = e^{-\tau^2\sigma^2/2 + i\omega_Z \tau}. \qquad (16)$$

Eq. 9 can be rewritten in terms of diagonal, $\rho_{\downarrow\downarrow}, \rho_{\uparrow\uparrow}$, and off-diagonal, $\rho_{\downarrow\uparrow}, \rho_{\uparrow\downarrow}$, components of the spin density matrix,

$$\dot{\rho}_{\uparrow\uparrow}(t) - \dot{\rho}_{\downarrow\downarrow}(t) = -\int_0^\infty d\tau \left( \rho_{\uparrow\uparrow}(t) - \rho_{\downarrow\downarrow}(t) \right) \times$$
$$\left\{ \sigma^2 \cos(\omega_Z + A_0/2)\tau + \sigma^2 \cos(\omega_Z - A_0/2)\tau + (A_0^2/2)\cos\omega_Z \tau \right\} e^{-\tau^2\sigma^2/2}, \qquad (17)$$

and

$$\dot{\rho}_{\uparrow\downarrow}(t) = \left[\dot{\rho}_{\downarrow\uparrow}(t)\right]^* = -\int_0^\infty d\tau\, \rho_{\uparrow\downarrow}(t) \left\{ \sigma^2 \cos(A_0/2)\tau + (A_0/2)^2 \right\} e^{i\omega_Z \tau - \tau^2\sigma^2/2}. \qquad (18)$$

Solving the above equations, we find that the off-diagonal components of the density matrix decay as $\exp(-t/T_2)$, where

$$\frac{1}{T_2} = \sqrt{\frac{\pi}{2\sigma^2}} \left( \frac{\sigma^2}{2} e^{-\frac{(\omega_Z + A_0/2)^2}{2\sigma^2}} + \frac{\sigma^2}{2} e^{-\frac{(\omega_Z - A_0/2)^2}{2\sigma^2}} + \left(\frac{A_0}{2}\right)^2 e^{-\frac{\omega_Z^2}{2\sigma^2}} \right). \qquad (19)$$

The electron spin polarization, $\rho_{\uparrow\uparrow} - \rho_{\downarrow\downarrow}$, decays according to $\exp(-t/T_1)$, where from Eq. 17 one obtains that $T_1 = T_2/2$. The three terms in Eq. 19 correspond to the channels of dissipation of the electron-spin phase. The first two terms describe donor electron spin-flips owing to its interaction with the $^{29}$Si nuclear spins. The third term arises due to donor electron spin-flips at the P-donor nucleus.

**Discussion and Summary.** The transverse relaxation rate, $1/T_2$, Eq. 19, is shown in Fig. 2 as a function of the external magnetic field. It has a peak of Gaussian shape at $H=0$, and another Gaussian peak at $H = A_0\hbar/2g\mu_b$. The intensity and width of the peaks are determined by the electron-spin hyperfine interaction constant, $A_0$, with the donor nucleus, and by the energy-scale $\sigma$, see Eqs. 14, 19. The latter parameter depends of the particular lattice arrangement of $^{29}$Si nuclei around the $^{31}$P impurity.

Indeed, thus far we have considered a single donor electron spin, and Eq. 19 described a decoherence process. In order to obtain a specific estimate, used for Fig. 2, we have assumed that one can average Eq. 9 over a statistical ensemble of spatially distributed $^{29}$Si nuclei surrounding the $^{31}$P donors. The average value, $\sigma^*$, then provides a representative measure of the individual spin decoherence. The quantity defined via

$$\left(\sigma^*\right)^2 = c \sum_l (A_l/2)^2, \qquad (20)$$

where $c$ is the concentration of the $^{29}$Si nuclei, and the summation is carried over all the lattice positions, can be obtained from experimental data on the inhomogeneously broadened ESR line of $^{31}$P donors.[27] For a Gaussian line,[29]

$$\sigma^* = \frac{g\mu_b \delta_{Si}}{2\hbar\sqrt{2\ln 2}}, \qquad (21)$$

where the width at the half intensity of the ESR line, $\delta_{Si} \approx 2.9\,\text{G}$, is obtained from experiment.[27]

The estimated value of $\sigma^*$ is thus $2.166 \times 10^7$ s$^{-1}$. The value of the hyperfine constant for the $^{31}$P donor nuclei is[25] $A_0 \approx 7.39 \times 10^8$ s$^{-1}$. Thus, the averaged value of $T_{1,2}$ for the donor electron spin in the Si:P system at low external magnetic field and low temperatures, can be as short as $T_1, T_2 \sim 10^{-7} - 10^{-9}$ s$^{-1}$. This indicates that the decoherence mechanism considered in this work is likely to be the dominant one at low magnetic fields.

In summary, we have considered a model of a shallow donor electron spin interacting with a nuclear spin bath. We have shown that at low temperatures such system relaxes to the state determined by the density matrix of the nuclear spin bath, on time scales or order $\sim 10^{-7}-10^{-9}$ s$^{-1}$ for low external magnetic fields. Within the approximation scheme used, the transition probabilities determining $T_1^{-1}$ and $T_2^{-1}$ have Gaussian dependence of the applied magnetic field.

**Acknowledgment.** This research was supported by the National Science Foundation, grants DMR-0121146 and ECS-0102500, and by the National Security Agency and Advanced Research and Development Activity under Army Research Office contract DAAD-19-99-1-0342.

**References**.

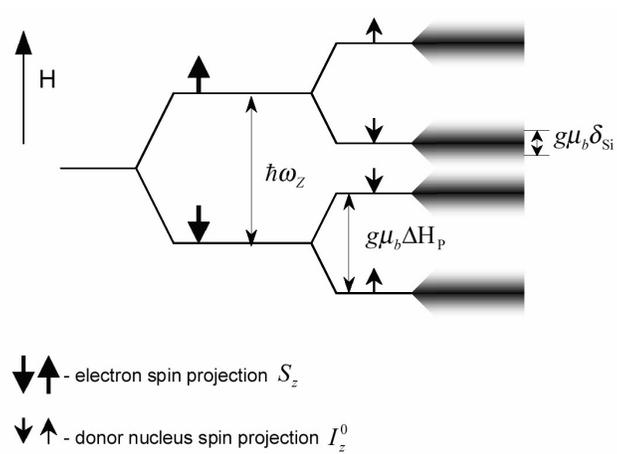

**Figure 1.** Energy level structure of a donor electron spin interacting with a system of $^{29}$Si nuclear spins via the contact hyperfine interaction; see Ref. 27.

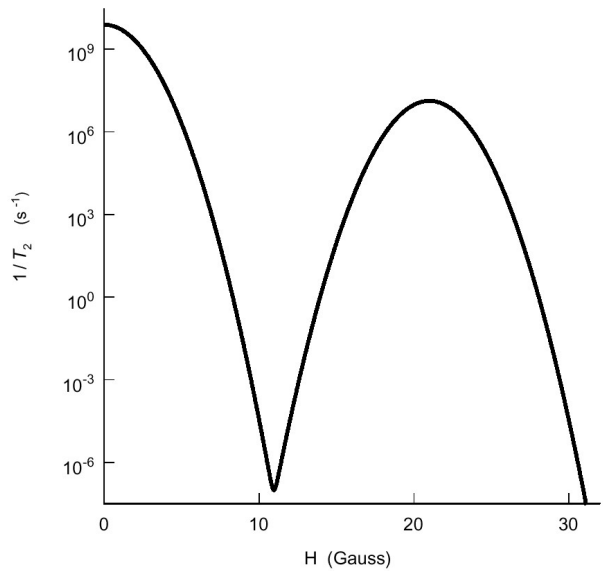

**Figure 2.** The transverse relaxation rate $1/T_2$, shown on a logarithmic scale, of a donor electron spin as a function of the external magnetic field.